\documentclass[12pt,a4paper,twoside,openright,prb]{article}
\usepackage{graphics}
\usepackage{caption2}
\newcommand{\be}{\begin{equation}}
\newcommand{\ee}{\end{equation}}
\newcommand{\ba}{\begin{eqnarray}}
\newcommand{\ea}{\end{eqnarray}}

\begin{document}
\baselineskip 0.65cm

\title{\bf Memory effects in classical and quantum mean-field disordered models}
\vskip 20pt

\author{ L. F. Cugliandolo$^{1,2}$, 
G. Lozano$^{3}$ and H. Lozza$^3$ \\
 {\footnotesize \it $^1$
Laboratoire de Physique Th{\'e}orique de l'{\'E}cole Normale
Sup{\'e}rieure,} \\
{\footnotesize \it 24 rue Lhomond, 75231 Paris
Cedex 05, France} \\
{\footnotesize \it $^2$ Laboratoire de Physique
Th{\'e}orique et Hautes {\'E}nergies, Jussieu, } \\ {\footnotesize \it
1er {\'e}tage, Tour 16, 4 Place Jussieu, 75252 Paris Cedex 05, France}
\\
{\footnotesize \it
$^3$ Departamento de F\'{\i}sica, FCEyN,
Universidad de Buenos Aires,
}
\\
{\footnotesize \it
 Pabell{\'o}n I, Ciudad Universitaria, 1428 Buenos Aires, Argentina}
}
\maketitle

\begin{abstract}
We apply the Kovacs experimental protocol to classical and quantum $p$-spin 
models. We show that these models have memory effects as those 
observed experimentally in super-cooled polymer melts. We discuss our
results in connection to other classical models that capture memory
effects. We propose that a similar protocol applied to quantum glassy 
systems might be useful to understand their dynamics.  
\end{abstract}

\newpage

\normalsize
\addtolength{\baselineskip}{0.2\baselineskip}

\section{Introduction}
\label{sec:intro}

The Kovacs memory effect was first reported by this author in the
60s~\cite{Kovacs}. It demonstrates that super-cooled liquids when
taken out of equilibrium have a very intricate dynamics that cannot be
predicted on the basis of the knowledge of the instantaneous value of
the state variables ($P,V,T$) right after the perturbation. More 
precisely,
Kovacs showed that the specific volume of a polymer melt in its
super-cooled liquid phase, has a rather non-trivial evolution that
depends on the thermal history of the sample~\cite{Kovacs,McKenna}.  
Non-trivial effects of temperature variations were also
observed in the evolution of (two-time) susceptibilities of dipolar
glasses~\cite{dipolar}, spin-glasses~\cite{spin-glasses} and many
other glassy systems~\cite{Struik,otherTshifts}.  
History-dependent phenomena in
granular compaction have been recently reported~\cite{Josserand}. 
In this case, the control parameter is the tapping strength and the
observable is the density.

The experimental setup involved in the Kovacs effect is the
following. In a first step, one quenches an equilibrated liquid with a
very fast cooling rate from a high temperature $T_0$ to a low
temperature $T_2$, at time $t=0$. One then follows the
subsequent evolution of the quantity of interest that in Kovacs'
experiments is the specific volume. In order to match what we shall
study in this paper, we describe the experiment using another {\it
one-time quantity}, the energy density, as the example. The energy
density relaxes in time and it slowly approaches an asymptotic value
that may fall out of the experimental time-window. The time-dependent
curve ${\mathcal E}^{(T_2)}(t)$ constitutes a reference and we use a
superscript $(T_2)$ to indicate that it has been obtained using the
first prescription, at fixed external temperature $T_2$. The
extrapolated asymptotic value defines ${\mathcal E}_{\sc as}(T_2)$. In a
second step, one first quenches the sample from the same high
temperature $T_0$ to a lower temperature $T_1$$ (<T_2)$ at time $t=0$,
waits until a carefully chosen time $t_1$ and then heats the sample to
the final temperature $T_2$. The value of $t_1$ is chosen such that
${\mathcal E}^{(T_1\to T_2)}(t_1^+)={\mathcal E}_{\sc as}(T_2)$, where the
plus sign indicates that the one-time quantity should match the
asymptotic value obtained with the first procedure right after heating
the sample from $T_1$ to $T_2$.  [In the experimental protocol the
need to use ${\mathcal E}^{(T_1\to T_2)}(t_1^+)={\mathcal E}_{\sc as}(T_2)$ is
due to the fact that when changing the temperature there is a trivial
response due to the thermal expansion of the local degrees of
freedom.]  Since the initial value of the energy density at the final
temperature $T_2$ is already the asymptotic one, one could have
expected that the energy remained constantly fixed to this value for
all subsequent times independently of the value of $T_1$ (apart
from 
very fast rearrangements decaying exponentially). However, Kovacs
demonstrated that after $t_1$ the energy-density, ${\mathcal E}^{(T_1\to
T_2)}(t)$, has a slow non-monotic dependence on time, first increasing
and then decreasing back to its initial and asymptotic value ${\cal
E}_{\sc as}(T_2)$:
\begin{equation}
{\mathcal E}^{(T_1\to T_2)}(t) =  {\mathcal E}_{\sc as}(T_2) + \Delta {\mathcal E}(t)
\; ,
\end{equation}
with the ``Kovacs hump'' $\Delta {\mathcal E}$ satisfying 
\begin{equation}
\Delta {\mathcal E}(t)>0
\; , 
\;\;\;\;\;\;\;\;\;\;\;
\Delta {\mathcal E}(t_1^+)=\lim_{t\to\infty} \Delta {\mathcal E}(t)=0
\; .
\label{condition-t1}
\end{equation} 
The form of the hump depends on the values $T_2$ and $T_1$ used.
Qualitatively, its height is an increasing function of  $T_2-T_1$
and the time at which the maximum is reached decreases when $T_2-T_1$ 
increases.

It is interesting to note that Kovacs' experiments have been performed
in the super-cooled liquid phase. While it would be no surprise to
find nonequilibrium effects in the glassy phase, the reason why one
also finds a nonequilibrium behaviour in the super-cooled liquid is
that the jump in the external temperature drives the system out of
equilibrium and the 
 relaxation occurs in a very long
time-scale. This experiment proves that the knowledge of 
the state variables, in Kovacs' experiments $P,V$ and $T$ 
at $t_1^+$, is not sufficient to determine the subsequent evolution
of the same quantities in the glassy and super-cooled liquid phases (if
the latter has been recently strongly perturbed).

Recently, several authors presented analytical and numerical studies
of this effect using a variety of models with glassy dynamics.  So
far, apart from phenomenological approaches~\cite{McKenna}, the Kovacs
effect has been analysed numerically with molecular dynamic
simulations of a molecular model of a fragile glass
former~\cite{Sciortino} and Montecarlo simulations of the $3d$
Edwards-Anderson spin-glass~\cite{Bebo}, and analytically within the
ferromagnetic Ising chain~\cite{Brawer}, the critical $2d$ {\sc xy}
model~\cite{Beho}, the trap model~\cite{Bertin}, domain
growth~\cite{Bertin}, $1d$ kinetically constrained spin
models of fragile and strong type~\cite{Buhot}, 
and the parking lot model of granular
matter~\cite{Tarjus}.  It has been suggested that the quantitative
analysis of the Kovacs effect may help distinguishing between
different glassy models and, perhaps, may also help identifying
spatial properties of glassy systems.  In this paper we show that the
main qualitative features of the classical Kovacs effect are captured
by the fully connected spherical $p$-spin disordered system~\cite{Crso}, 
a model with no
spatial structure but with a slow dynamics leading to very slow
relaxations close to the transition to and in the
glassy phase~\cite{Cuku93,Leto}.  This `negative' result, as far as
what can be deduced about {\it spatial rearrangements} from the Kovacs
effect, is reminiscent of the discussion~\cite{JLuis} on the
interpretation of hole burning experiments~\cite{hole-burning}.  We
also discuss the scaling laws that describe the behaviour of the hump
and compare them to what found in other glassy models.

On the other hand, the analysis of quantum glassy systems is now
starting to call the attention of experimentalists and
theoreticians. The slow, history-dependent 
relaxation  of a dipolar 
quantum system in its glassy phase has been 
reported~\cite{aeppli}.  The sample that entered the
glassy phase following a quantum route (changing the strength of quantum 
fluctuations at fixed low temperature) is always in advance with
respect to the one that arrived at the same point in parameter space
following a classical path (keeping the strength of quantum fluctuations
fixed and cooling the system). This has been demonstrated by the fact
that the time-dependent dielectric constant of the quantum-cooled
sample is closer to its asymptotic value at any finite, experimental
time. Memory effects were also observed in glasses at ultralow 
temperatures~\cite{Osheroff} and the electron glass~\cite{Zvi}.

A variant of the Kovacs procedure where the control parameter is the
strength of quantum fluctuations can be easily envisaged.  The
question then arises as to whether a hump appears and which is
its structure, scaling form, etc.  We address this question using a
quantum extension of the $p$-spin model introduced and studied in
\cite{Culo,Cugrlolo} (see also~\cite{others}).

In short, in this paper we show that simple disordered mean-field
models capture the phenomenology of the Kovacs effect. With this aim,
we analyse the nonequilibrium relaxation of the spherical $p$-spin
disordered model in its classical~\cite{Cuku93,Leto} and
quantum~\cite{Culo} versions, see Sect.~\ref{sect:model} for their
definitions.  First, we reproduce the classical setup using
temperature as the control parameter and we discuss the results in comparison 
with previous explanations of the same effect (Sect.~\ref{sect:classical}).
Second, we switch on quantum fluctuations and use their strength as
the control parameter (Sect.~\ref{sect:quantum}). In both cases we
follow the evolution in time of the potential energy-density of the system. We
present our conclusions in Sect.~\ref{sect:conc}.

\section{The spherical p-spin model}
\label{sect:model}

The spherical $p$ spin model is defined by the Hamiltonian~\cite{Crso}
\begin{equation}
H_J[{\vec S}] = - \sum_{\langle i_1 i_2 \dots i_p\rangle} 
J_{i_1 i_2 \dots i_p} s_{i_1} s_{i_2} \dots s_{i_p}  
\; .
\label{pspin-hamil-class}
\end{equation}
The spins $s_i$ are continuous variables $s_i \in (-\infty,\infty)$
forced to satisfy the global spherical constraint $\sum_{i=1}^N s_i=N$
with $N$ their total number in the sample.  
The exchange interaction  $J_{i_1 i_2
\dots i_p}$ are quenched random variables drawn from a Gaussian
distribution with average $[J_{i_1\dots i_p}]=0$ and variance
$[J_{i_1\dots i_p}^2]={\tilde J}^2 p!/(2N^{p-1})$. We henceforth 
use square brackets to indicate the average over
disorder. The interactions occur between all groups of $p$ spins
in the sample. The model is then {\it fully-connected} and mean-field
in character. The scaling of the variance $[J_{i_1\dots i_p}^2]$ with
$N$ has been chosen so as to ensure a good thermodynamic limit.  The
parameter $p$ takes integer values, $p \geq 2$.  Mixed models with 
a Hamiltonian with two terms of the form (\ref{pspin-hamil-class}) with 
$p=2$ and $p=4$~\cite{Gotze} are also of interest~\cite{Leto}.

The classical dynamics is determined by the Langevin equation 
\begin{equation}
\dot s_i(t) = - \frac{\partial E(t)}{\partial s_i(t)} - 
\mu(t) s_i(t) + \xi_i(t)
\end{equation}
where $E(t)$ is the total energy, 
$\mu(t)$ is the Lagrange multiplier enforcing the spherical constraint,
and $\xi_i(t)$ is a Gaussian white-noise with zero mean and correlations
\begin{equation}
\langle \xi_i(t) \xi_j(t') \rangle = 2 k_B T \delta_{ij} \delta(t-t')
\; .
\end{equation}
We assume that the system is prepared at $t=0$ with an infinitely
fast quench from $T_0\to\infty$ to the initial temperature $T_1$. 
The dynamic equations for the macroscopic order parameters 
are derived using standard functional techniques~\cite{Leto}. We discuss them 
below, as the classical limit of the quantum extension of the same model.
The models 
with $p=2$, 
$p\geq 3$ 
 have different dynamics; 
the former yields a mean-field 
description of simple domain growth, while the latter  case is related to 
the schematic mode-coupling theory ({\sc mct}) of super-cooled liquids and 
glasses. 
These models and their generalizations are reviewed in detail in~\cite{Leto}.

Quantum fluctuations can be introduced~\cite{Culo} 
by upgrading the spins $s_i$ to 
coordinate-like operators, adding a `kinetic term' 
\begin{equation}
\frac{1}{2M} \, \sum_{i=1}^N \hat \Pi_i^2
\end{equation}
to the Hamiltonian (\ref{pspin-hamil-class}), and imposing canonical 
commutation relations
\begin{equation}
[\hat \Pi_i, \hat s_j] = i \hbar \, \delta_{ij}
\; ,
\;\;\;\;\;\;\;\;
[\hat s_i, \hat s_j] = 0
\; ,
\;\;\;\;\;\;\;\;
[\hat \Pi_i, \hat \Pi_j] = 0
\; .
\end{equation}
At time $t=0$ we set the model in contact with an 
Ohmic bath of quantum harmonic oscillators 
\begin{eqnarray}
H_b &=& 
\sum_{l=1}^{\tilde{N}} \frac1{2m_l} \hat{p}_{l}^2 +
\sum_{l=1}^{\tilde{N}} \frac12 m_{l} \omega_{l}^2 \hat{x}_{l}^2 \; ,
\\
H_{sb} &=&
-\sum_{i=1}^N  {\hat s}_i^{z} \sum_{l=1}^{\tilde{N}}
c_{i l} \hat{x}_{l}
\; ,
\end{eqnarray}
where, for simplicity, we considered a bilinear coupling, $H_{sb}$.
(Note that for the spherical problem it is not necessary to introduce
a counterterm). We assume that this 
environment has a well-defined temperature and that it is not 
modified by the interaction with the system. The initial 
density matrix is factorised and we choose random initial conditions 
for the system. After integrating out the bath degrees of freedom, and using 
the fully-connected character of (\ref{pspin-hamil-class}), one 
arrives at a dynamic generating functional from which one derives 
exact dynamic equations for the 
macroscopic two-time dependent order 
parameters
\begin{equation}
C(t,t') = \frac{1}{N} 
\sum_{i=1}^N [\langle \{\hat s_i(t), \hat s_i(t')\}  \rangle]
\; ,
\;\;\;\;\;\;\;\;\;\;\;\;
R(t,t') = \frac{1}{N} \sum_{i=1}^N 
\left. \frac{[\langle \hat s_i(t)\rangle]}{h_i(t')} \right|_{h=0}
\; .
\end{equation}
$h$ is an infinitesimal field that couples linearly to 
the spins, modifying the Hamiltonian as $H\to H- \sum_{i=1} h_i \hat s_i$. 
$C$ is the symmetrized correlation function and $R$ is the linear response 
of the system.

The dynamic equations take the Schwinger-Dyson form
\begin{eqnarray}
&&
[M\partial^2_t  + \mu(t)] \, 
R(t,t') 
= \delta(t-t') + 
\int_{t'}^t dt'' \, \Sigma(t,t'') R(t'',t')
\; ,
\label{schwingerR}
\\
&&
[M\partial^2_t  + \mu(t)] \, 
C(t,t') 
=
\int_{0}^t dt'' \, \Sigma(t,t'') C(t'',t') +
\int_{0}^{t'} dt'' \, 
D(t,t'') R(t',t'')
\; ,
\label{schwingerC}
\end{eqnarray}
with the self-energy $\Sigma$ and the vertex $D$
given by 
\begin{eqnarray}
\Sigma(t,t'') &=& - 4 \eta(t-t') + \sigma(t,t')
\; ,
\\
D(t,t'') &=& 2\hbar \nu(t-t') + d(t,t') 
\; .
\end{eqnarray}
The first contributions originate in the interaction with the 
Ohmic bath of spectral density~\cite{Culo,Cugrlolo}
\begin{equation}
I(\omega) = \frac{4\gamma}{\pi} \omega e^{-\omega/\Lambda}
\, \theta(\omega)
\; ,
\end{equation}
where $\Lambda$ is a cut-off included to avoid divergences. 
The kernels $\eta$ and $\nu$ are functions of the 
time-difference $\tau\equiv t-t'$ and they are  given by 
\begin{eqnarray}
\eta(\tau) &=& 
- \frac{4\gamma \Lambda^2}{\pi} 
\frac{2\Lambda\tau }{(1+(\Lambda\tau)^2)^2
}
\; ,
\\
\nu(\tau) &=& \frac{4\gamma}{\pi} 
\int_0^\infty d\omega \; \omega e^{-\omega/\Lambda} 
\mbox{coth} \left( \frac{\beta\hbar\omega}{2} \right) \, \cos(\omega\tau)
\; .
\end{eqnarray}
The second contributions, $\sigma$ and $d$,
are due to the interactions in the system and 
read 
\begin{eqnarray}
\sigma(t,t') &\equiv&
-\frac{p {\tilde J}^2 }{\hbar} 
\mbox{Im} \left[ C(t,t')-\frac{i \hbar}{2} R(t,t') \right]^{p-1} \; ,
\label{kernelSigma}
\\ 
d(t,t') &\equiv &
\frac{p {\tilde J}^2}{2}
\mbox{Re}\left[ C(t,t')-\frac{i\hbar}{2}( R(t,t')+ R(t',t)) \right]^{p-1}
\; . 
\label{kernelD}
\end{eqnarray}
An integral equation that fixes the Lagrange multiplier $\mu(t)$ supplements
these equations and it is derived from the requirement $C(t,t)=1$ for all 
times~\cite{Culo}.

The classical limit is easily obtained by 
neglecting the kinetic term 
and by  taking the limit $\hbar\to 0$. Indeed, in this case, the effect of
the coupling to the bath reduces to the usual contributions originating in the
friction and noise terms of the Langevin equation and involving a 
first-order time-derivative of the correlation and response when 
$\Lambda$ is further taken to infinity.

In the quantum case, three contributions to the total energy density
can be identified: the kinetic part, the potential part and the interaction 
with the bath. In the following we focus on the averaged potential energy
density
\begin{equation}
{\mathcal E}(t)
\equiv 
N^{-1} [ \langle H_J[{\vec S}] \rangle ] 
= 
- \frac{1}{N}  \left[
\left\langle 
\sum_{\langle i_1 i_2 \dots i_p\rangle} 
J_{i_1 i_2 \dots i_p} \hat s_{i_1}(t) \hat s_{i_2}(t) \dots \hat s_{i_p}(t) 
\right\rangle
\right] \; .
\end{equation}
With a simple calculation one finds 
\begin{equation}
{\mathcal E}(t) = 
\int_0^\infty dt' \; \left[ \sigma(t,t') C(t,t') + d(t,t') R(t,t') \right]
\end{equation}
that in the classical case becomes
\begin{equation}
{\mathcal E}(t)= - \frac{\tilde J^2 p}{2} \int_0^t dt' \; C^{p-1}(t,t') R(t,t')
\; .
\end{equation}

\section{Classical case}
\label{sect:classical}

In this Section we analyze the classical problem with $p=3$.  After recalling the
value of the asymptotic energy-density for the isothermal relaxation,
we solve numerically the dynamic equations for $C$ and $R$ following
Kovacs' protocol. We analyse the Kovacs hump and we obtain its scaling with 
time and temperature in the three regimes of short, intermediate and long times. 

\subsection{Analytic results}

In the high-temperature phase the Fluctuation Dissipation Theorem ({\sc fdt}) 
implies
\begin{eqnarray}
{\mathcal E}_{as}(T) \equiv 
- \frac{p}{2} 
\lim_{t\to\infty} \int_{0}^{t} dt' \; C^{p-1}(t,t') 
\frac{1}{T} \frac{\partial C(t,t')}{\partial t'}
=- \frac{1}{2T} 
\; ,
\end{eqnarray}
$k_B=\tilde J = 1$ henceforth. We shall focus on the $p=3$ model
and use $T_2=0.75$ as the classical reference case for which 
${\mathcal E}_{as} (T_2) \approx -0.67$ (see Fig.~\ref{fig:energy-density}).

This model undergoes a  dynamic transition from an equilibrium (paramagnetic)
to a nonequilibrium (glassy) phase at 
\begin{equation}
T_d=\sqrt{\frac{p (p-2)^{p-2}}{2 (p-1)^{p-1}}}
\; .
\label{eq:criticalT}
\end{equation}
For $p=3$, $T_d \approx 0.61$ and the asymptotic energy-density at the
dynamic critical temperature takes the value ${\mathcal E}_{as}(T_d) \approx
- 0.82$.

In the low-temperature phase the solution 
of equations (\ref{schwingerR})-(\ref{kernelD}) 
involves a modification of the {\sc fdt}~\cite{Cuku93,Leto},
$R_s(t,t') = T_{eff}^{-1} \partial_{t'} C_s(t,t') \theta(t-t')$, for the slow
part of the relaxation. The effective temperature~\cite{Cukupe} 
 is given by $T_{eff}=(q_{ea} T)/[(p-2)(1-q_{ea})]$ 
and the energy-density reads
\begin{eqnarray}
{\mathcal E}_{as}(T) = 
- \frac{1}{2T} 
\left[(p-2)(1-q_{ea})q_{ea}^{p-1} + (1-q_{ea}^p)\right]
\label{eq:energy}
\end{eqnarray}
with the Edwards-Anderson order parameter, $q_{ea}$, determined by
\begin{eqnarray}
\frac{p(p-1)}{2} q_{ ea}^{p-2} (1-q_{ea})^2 = T^2
\; .
\label{eq:qea}
\end{eqnarray}
At the dynamic transition $T_{eff}=T$, $q_d=(p-2)/(p-1)$ and $T_d$ is
given by Eq.~(\ref{eq:criticalT}).
The numerical solution of Eqs.~(\ref{eq:energy}) and (\ref{eq:qea}) 
yields the value of the 
asymptotic energy-density in the glassy phase.

\subsection{Numerical results}

\begin{figure}[t!] \begin{center}
\resizebox{11cm}{!}{\includegraphics*[4.5cm,12cm][17.5cm,20cm]
{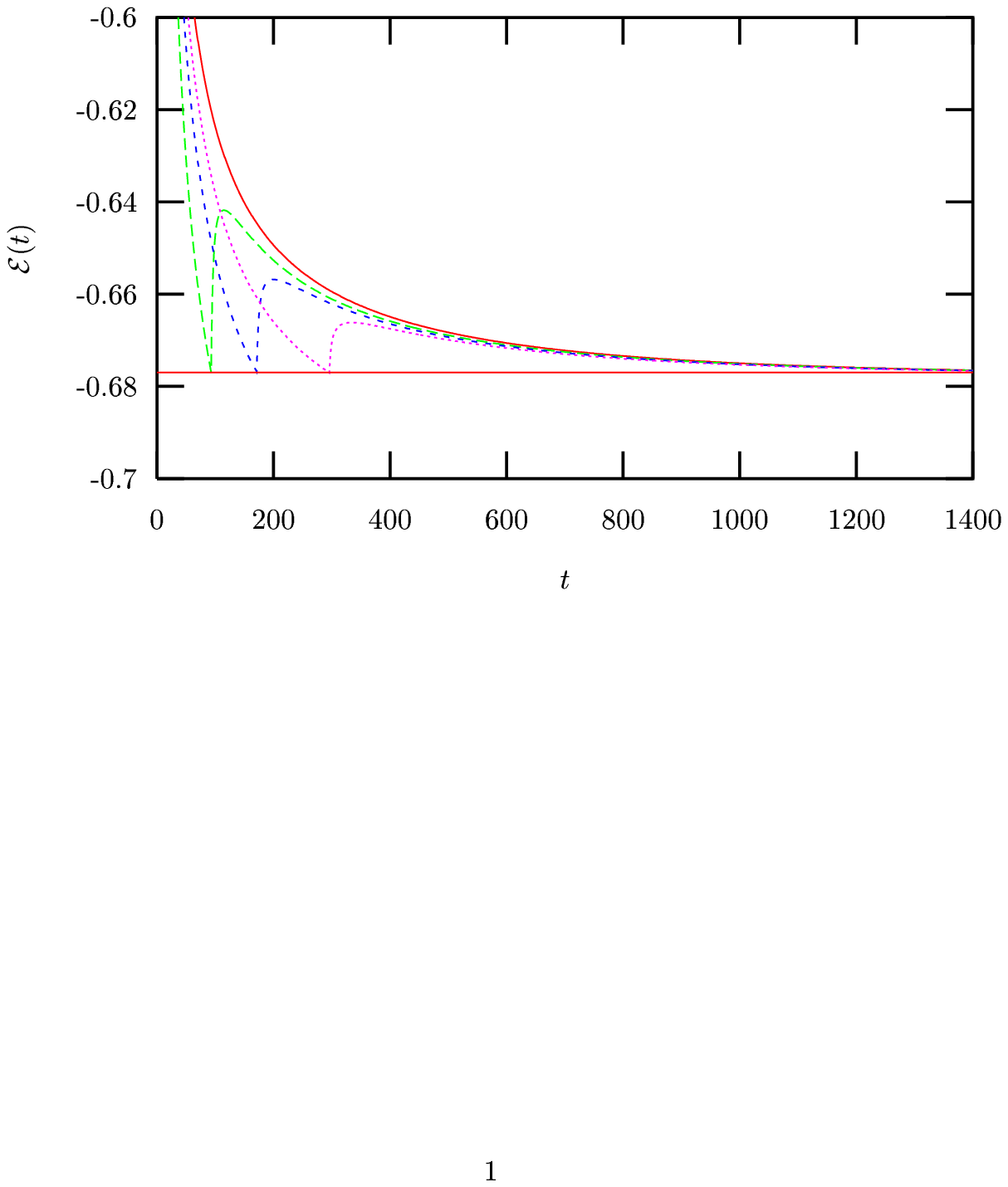}}
\setcaptionwidth{13cm}
\caption{\footnotesize The evolution in time of the energy-density  ${\mathcal
E}(t)$ in the classical $p=3$ spin model close and above the
paramagnetic -- spin-glass transition.  The solid line is
${\mathcal E}^{(T_2)}(t)$ and has been obtained using a rapid
quench to $T_2=0.75$. The other curves show  ${\mathcal E}^{(T_1 \to
T_2)}(t)$, {\it i.e.} the result of  having performed a temperature
jump from $T_1=0.65, 0.7, 0.725$  to $T_2$ at  $t_1=92$, $170$,
$295$ time-steps, respectively. The asymptotic  value ${\mathcal
E}_{as}(T_2)\approx -0.677$ is shown with a  horizontal line.}
\label{fig:energy-density}
\end{center} \end{figure}

The effect of temperature variations on the dynamics of the 
$p$ spin models in the glassy phase was studied in a couple of papers. The effect of 
small amplitude temperature cycles 
 on the nonequilibrium relaxation of the 
$p=2$ model were discussed in \cite{Cude} while 
their influence on the dynamics of the $p=3$
 was analyzed in \cite{theorTjumps}.

We show here that the $p$ spin model with $p\geq 2$ captures a similar phenomenology in the 
sense that a hump with a slow relaxation is obtained when the Kovacs'
protocol is applied. The model with $p=2$ is related to ferromagnetic 
domain growth, as described by the $O(N)$ model in $d=3$,  
and our results are intimately related to the ones 
in \cite{Brawer,Bertin}. 
In the following we present the data for $p=3$ (the schematic 
mode-coupling-theory~\cite{Leto,Gotze}) only.

Figure~\ref{fig:energy-density} shows the time-evolution of the 
energy-density in the $p=3$ classical model using different temperature 
jumps chosen according to Kovacs' rule. The solid line has been obtained 
using a quench to $T_2=0.75$. Note that since $T_2>T_d$ the 
quench is done within the paramagnetic 
phase in which the equilibration time is finite.
The asymptotic value approached with the algorithm is ${\mathcal E}_{as}(T_2) 
\approx -0.677$. (The analytic prediction is
${\mathcal E}_{as}(T_2) \approx -0.667$. Here and in what follows we use a 
time step $\delta=0.05$ to numerically solve the integro-differential 
equations.) The other curves include a temperature 
jump from $T_1$ to $T_2$ at $t_1$ when 
${\mathcal E}^{(T_1 \to T_2)}(t_1)={\mathcal E}_{as}(T_2)$.
The values of the parameters $T_1$ and $t_1$ are 
given in the caption. As observed by Kovacs
experimentally~\cite{Kovacs} and for a variety of 
models~\cite{Sciortino}~--~\cite{Tarjus}, the evolution of 
the energy-density is non-monotonic. Since the time $t_1$ is chosen so as 
to have 
${\mathcal E}^{T_1 \to T_2}(t_1^+)={\mathcal E}_{as}(T_2)\approx -0.677$, after the 
temperature jump the energy density already has the expected asymptotic 
value. However, it evolves during approximately three decades before reaching 
the reference curve, first increasing towards a 
maximum and then decreasing towards ${\mathcal E}_{as}(T_2)$. 

\begin{figure} \begin{center}
\resizebox{11cm}{!}{\includegraphics*[4.5cm,12cm][17.5cm,20cm]
{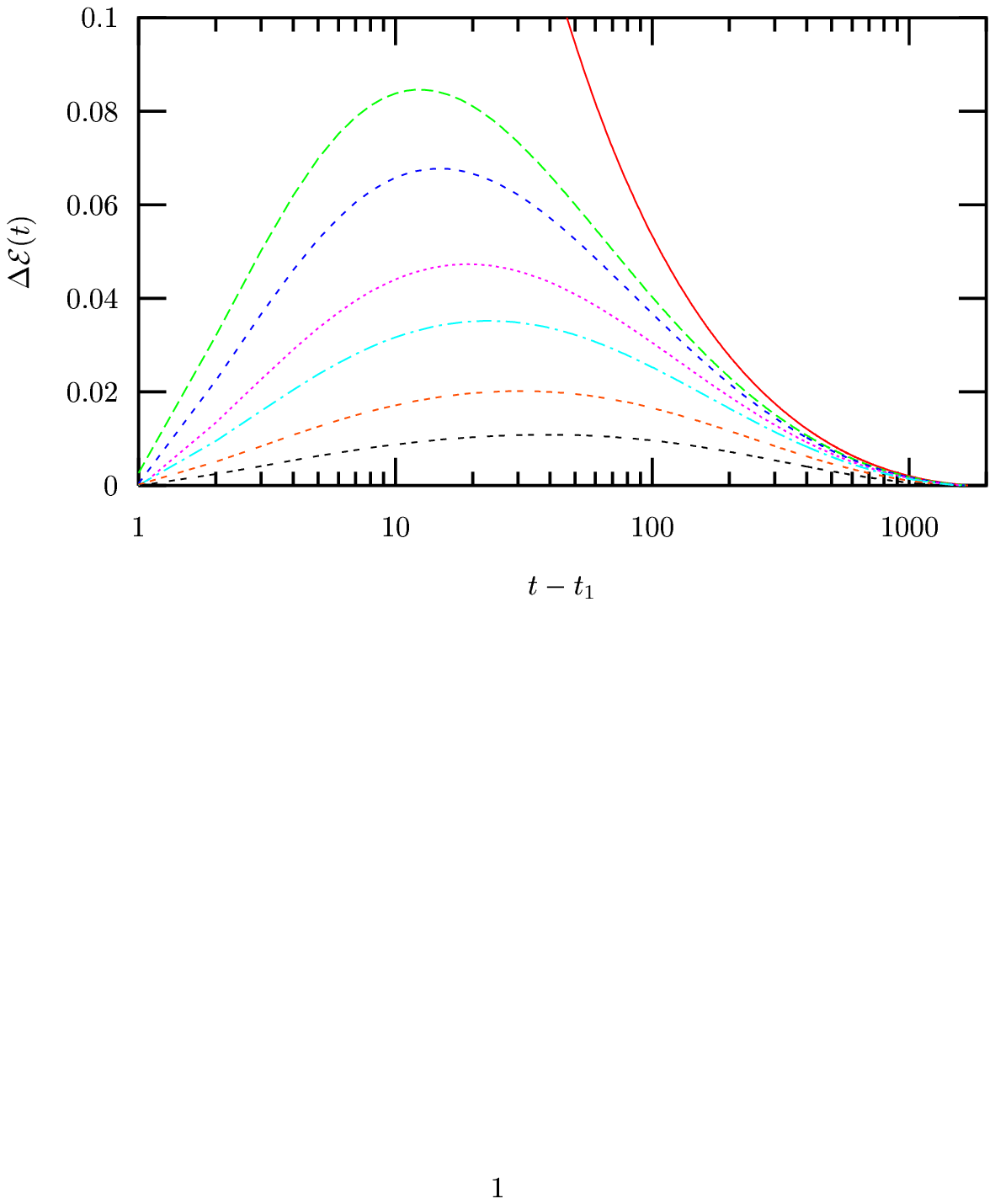}}
\setcaptionwidth{13cm} \caption{\footnotesize The hump, $\Delta {\mathcal E}(t)$, 
as a funtion of the time-difference, $t-t_1$, in the classical 
$p$ spin model with $p=3$ close and above its paramagnetic -- spin-glass 
transition. The different curves correspond to the 
intermediate temperatures $T_1=0.4$, $0.5$, $0.6$,
$0.65$, $0.7$, $0.725$ (from top to bottom) and the reference curve 
has been obtained for $T_2=0.75$.}
\label{fig:Khump-T}
\end{center} \end{figure}

\begin{figure} \begin{center}
\resizebox{12cm}{!}{\includegraphics*[4.5cm,12cm][17.5cm,20cm]
{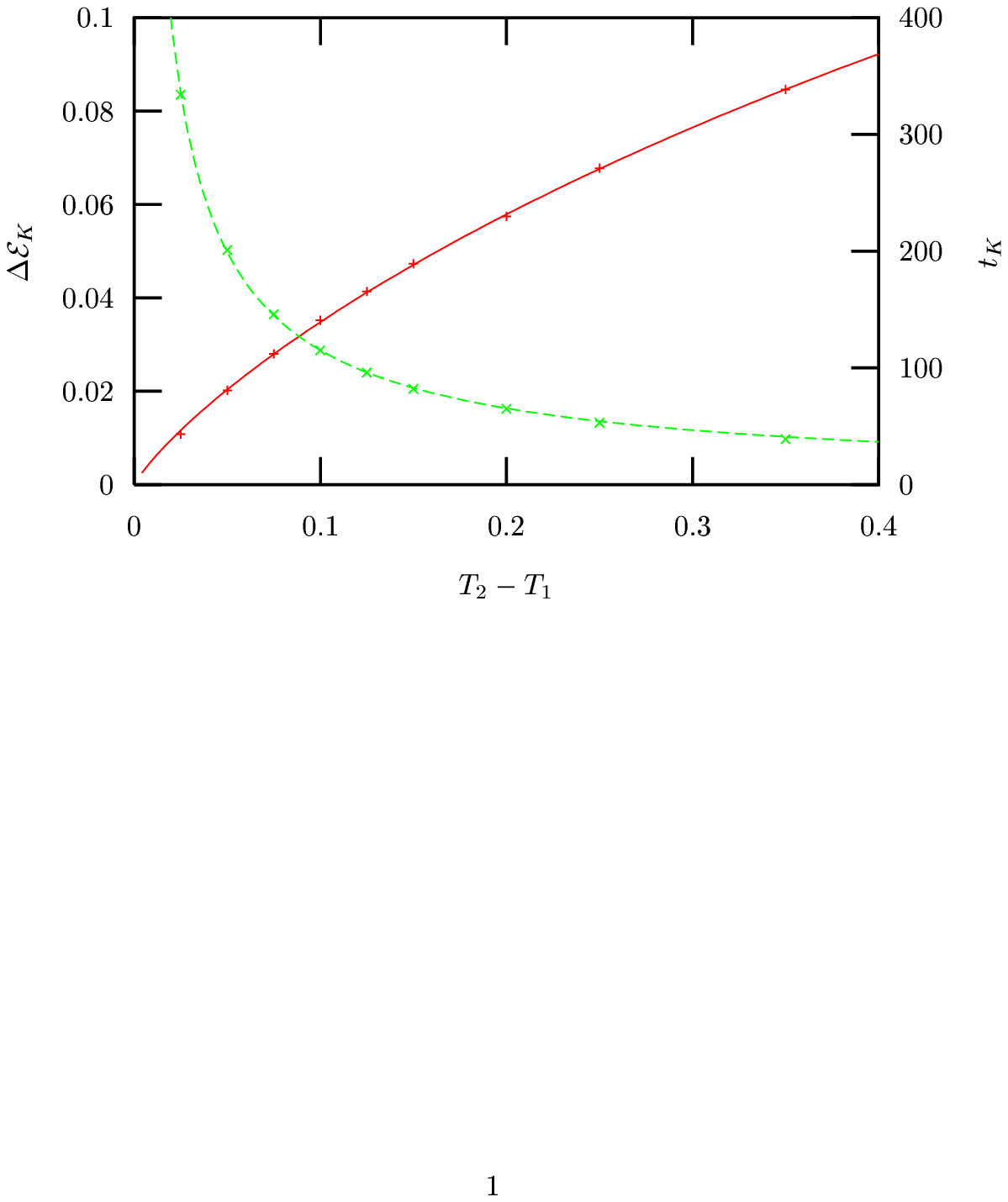}}
\setcaptionwidth{13cm} \caption{\footnotesize The maximum height of the hump, $\Delta {\mathcal E}_K(t)=
\; max \; \Delta {\mathcal E}$ 
(red curve, $+$) and its position $t_K\equiv t|_{max}$ 
(green curve, $\times$)
as a funtion of the difference in temperature $T_2 -T_1$ in the classical 
$p$ spin model with $p=3$ at $T_2=0.75$. The lines are fits to 
power laws with logarithmic corrections, see Eqs.~(\ref{fit:Emax}) 
and (\ref{fit:tK}),
with parameters $a=0.153$ and $b=0.556$ for the height of the maximum and
$c=16.73$ and $d=-0.135$ for its position.}
\label{fig:max-time}
\vspace{0.35cm}
\resizebox{11cm}{!}{\includegraphics*[4.5cm,12cm][17.5cm,20cm]
{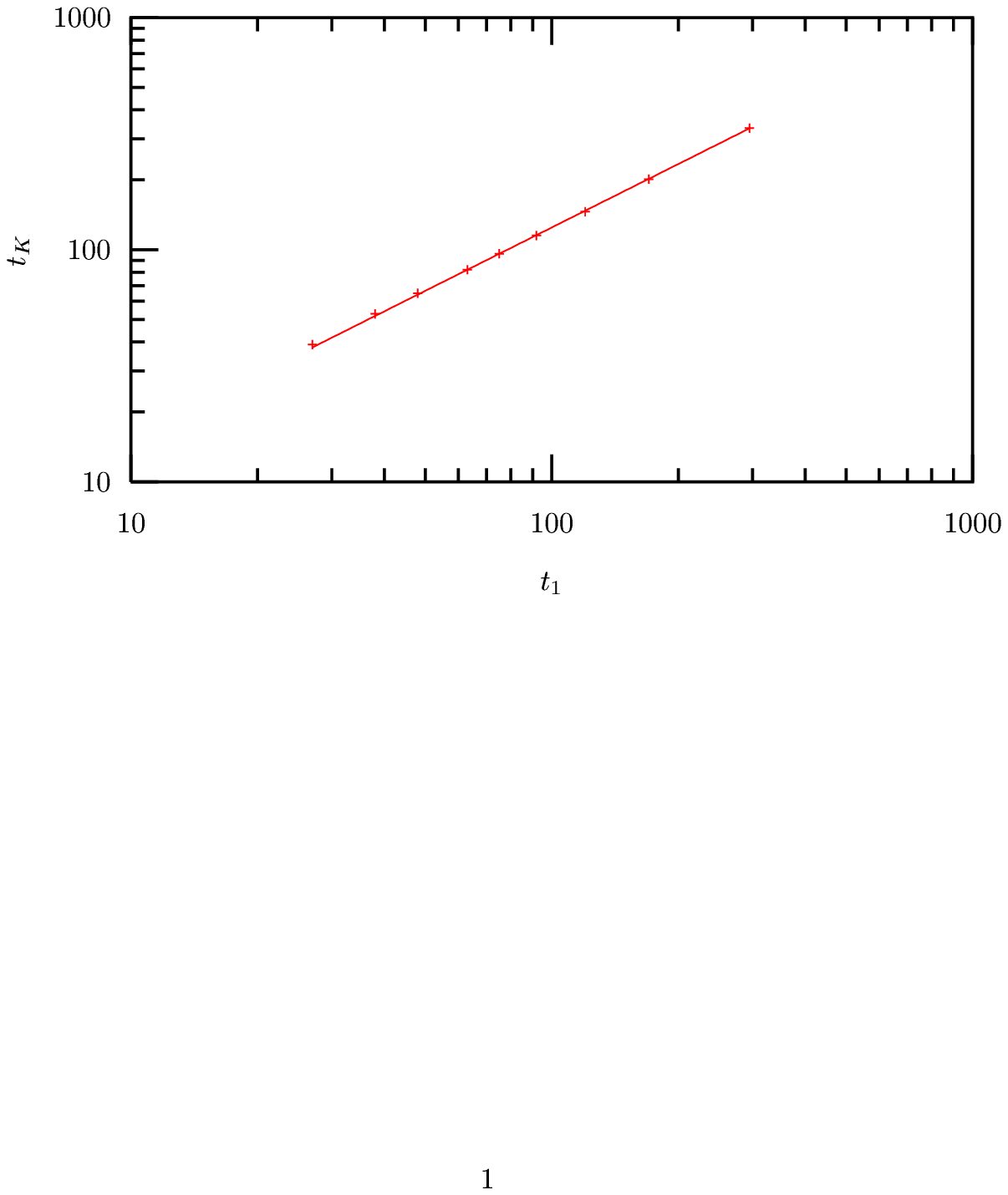}}
\setcaptionwidth{13cm} \caption{\footnotesize The position of the maximum in the hump, 
$t_K\equiv t|_{max}$, as a function of the time when the 
jump in temperature is imposed, $t_1$. The line is the power-law
relation $t_K = a \, t_1^b$ with $a=1.901$ and $b=0.908$.}
\label{fig:tK-t1}
\end{center} \end{figure}

The qualitative features of this non-monotonic behaviour are better
described by analysing the hump $\Delta {\mathcal E}$.
Figure~\ref{fig:Khump-T} shows its dependence with the temperature
difference $T_2-T_1$. As in the experimental data, the height of the
hump, $\Delta {\mathcal E}_K\equiv \; max \; \Delta
{\mathcal E}$, is an increasing function of $x\equiv T_2-T_1$, 
while the position of the
maximum, $t_K\equiv t|_{max}$, is a decreasing function of the same parameter.
 In Fig.~\ref{fig:max-time} we show that these quantities are well described 
by functions of the type  
\begin{eqnarray}
\Delta {\mathcal E}_K(x) &=& a \, x \, (1-b \ln x) 
\label{fit:Emax}
\\
t_K(x) &=& \frac{c}{x} \, (1- d \ln x)   
\label{fit:tK}
\end{eqnarray}
with $a,b,c,d$ fitting parameters.

We also show in Fig.~\ref{fig:tK-t1} that $t_K$, the position of the maximum, can be fitted as $t_K=a t_1^b$, where $t_1$ is the time when the temperature jump has been applied, and $a$ and $b$ are constants.

\vspace{0.35cm}

Three time-regimes can be identified in the  hump: short times well before the 
maximum is reached, intermediate times around the maximum and long-times when
the hump approaches its asymptotic vanishing value. The time-temperature 
dependence in the three regimes can be summarized as follows.

\vspace{0.25cm}
\noindent\underline{Short times}
\vspace{0.25cm}

\begin{figure} \begin{center}
\resizebox{11cm}{!}{\includegraphics*[4.5cm,12cm][17.5cm,20cm]{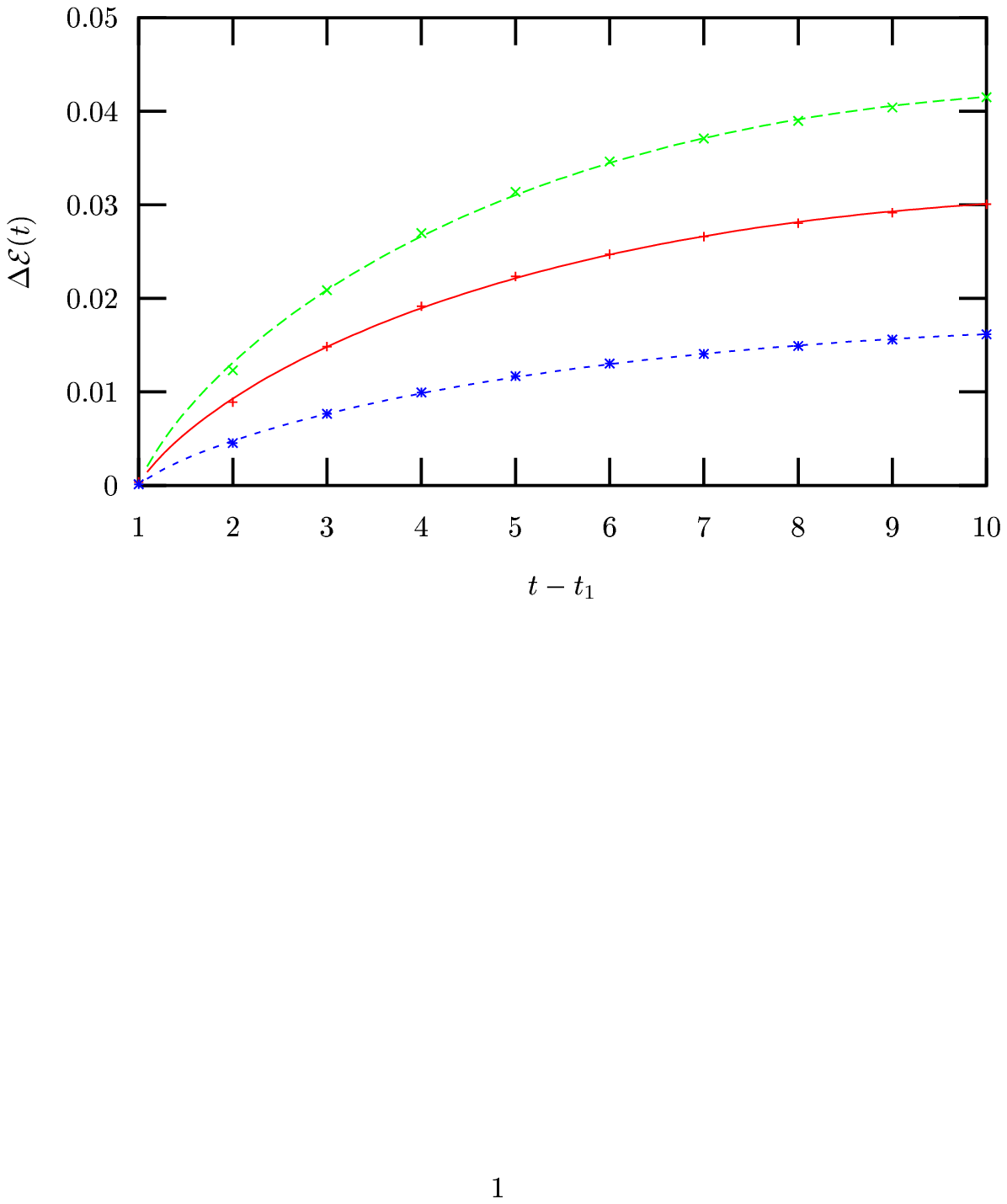}}
\setcaptionwidth{13cm} \caption{\footnotesize The short time fitting of the 
data for three values of the lower
temperature, $T_1=0.6, 0.65, 0.7$, using Eq.~(\ref{eq:scaling-shortt2}).}
\label{fig:short-times2}
\vspace{0.35cm}
\resizebox{11cm}{!}{\includegraphics*[4.5cm,12cm][17.5cm,20cm]{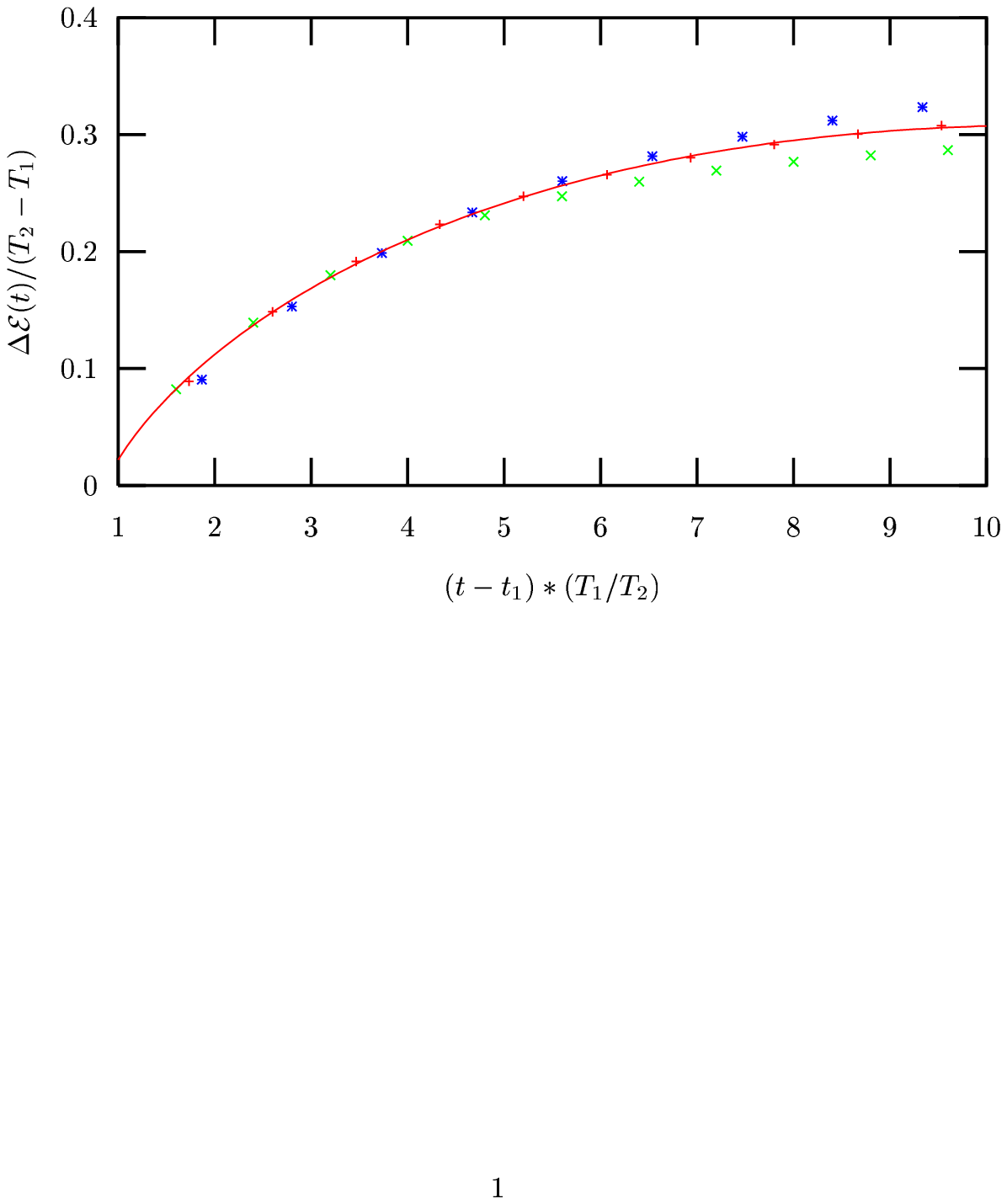}}
\setcaptionwidth{13cm} \caption{\footnotesize The short time scaling of the 
data for three values of the lower
temperature, $T_1=0.6, 0.65, 0.7$, using Eq.~(\ref{eq:scaling-shortt}). 
The solid line is a fit of the rescaled data for $T_1=0.65$ using  
Eq.~(\ref{eq:scaling-shortt2}) with $a=0.0623$ and $b=0.251$.}
\label{fig:short-times1}
\end{center} \end{figure}

In the short time regimen, a linear function with logarithmic corrections \cite{Bertin}
\begin{eqnarray}
{\cal F}_s(x) = a \, x \, (1-b\ln x)
\label{eq:scaling-shortt2}
\end{eqnarray}
fits the data with great precition. Indeed, in Fig.~\ref{fig:short-times2} we show together the data for $T_1=0.6, \, 0.65, \, 0.7$ and $t_1=63, \, 92, \, 170$ respectively and the fits by ${\cal F}_s(x)$.

The data can also be scaled using
\begin{eqnarray}
\Delta {\mathcal E}(t) &=& (T_2-T_1) 
{\cal F}_s \left((t-t_1) \frac{T_1}{T_2} \right)
\; .
\label{eq:scaling-shortt}
\end{eqnarray}
An accurate description of the rescaled data  is achieved
taking $a=0.0623$ and $b=0.251$ for ${\cal F}_s(x)$, 
see Fig.~\ref{fig:short-times1}.

\vspace{0.25cm}
\noindent
\underline{Intermediate times}
\vspace{0.25cm}

\begin{figure} \begin{center}
\resizebox{11cm}{!}{\includegraphics*[4.5cm,12cm][17.5cm,20cm]{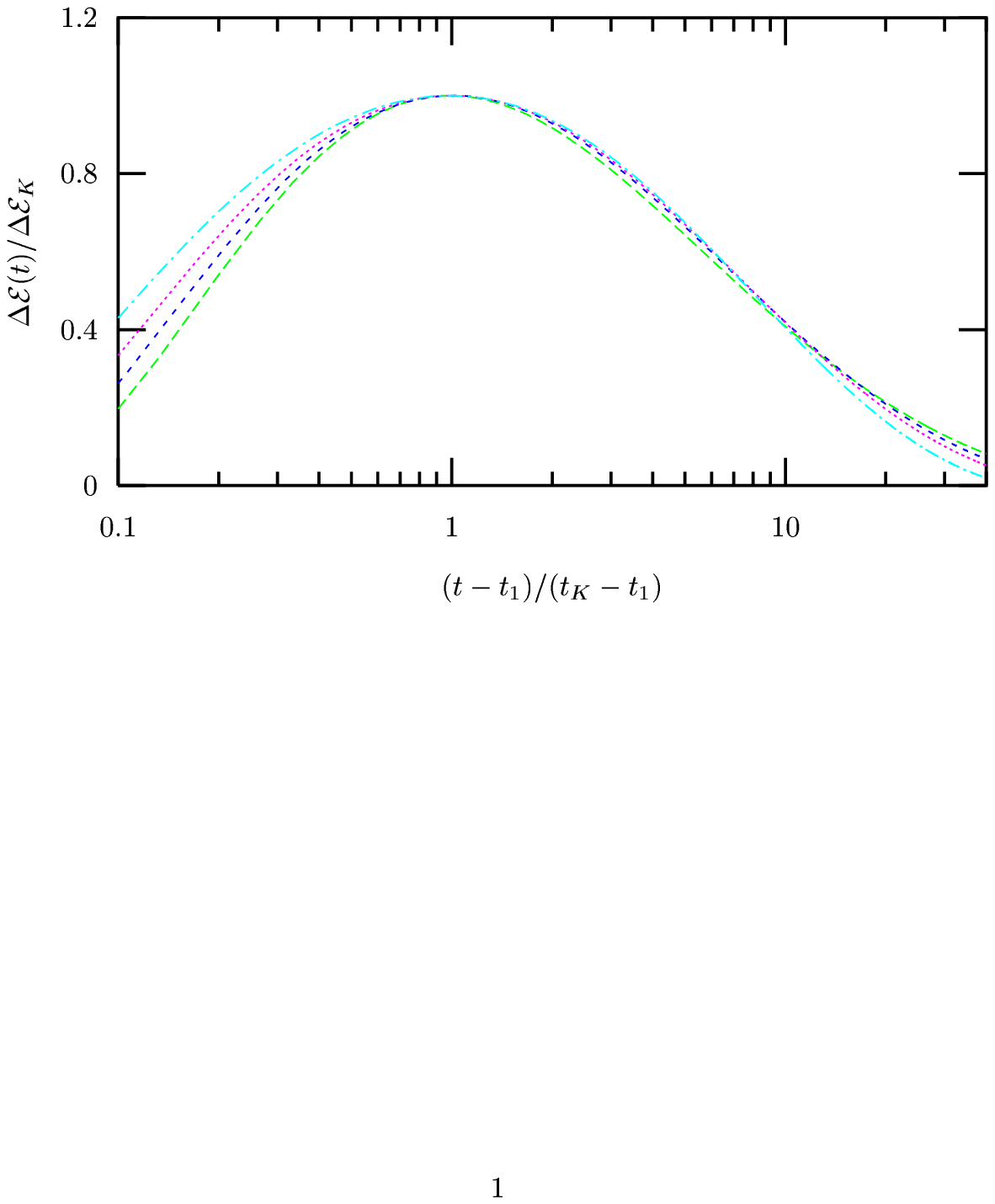}}
\setcaptionwidth{13cm} \caption{\footnotesize The scaling for intermediate times with $T_1=0.5,0.6,0.65,0.7$, see the text for an explanation.}
\label{fig:scaling}
\vspace{0.35cm}
\resizebox{11cm}{!}{\includegraphics*[4.5cm,12cm][17.5cm,20cm]{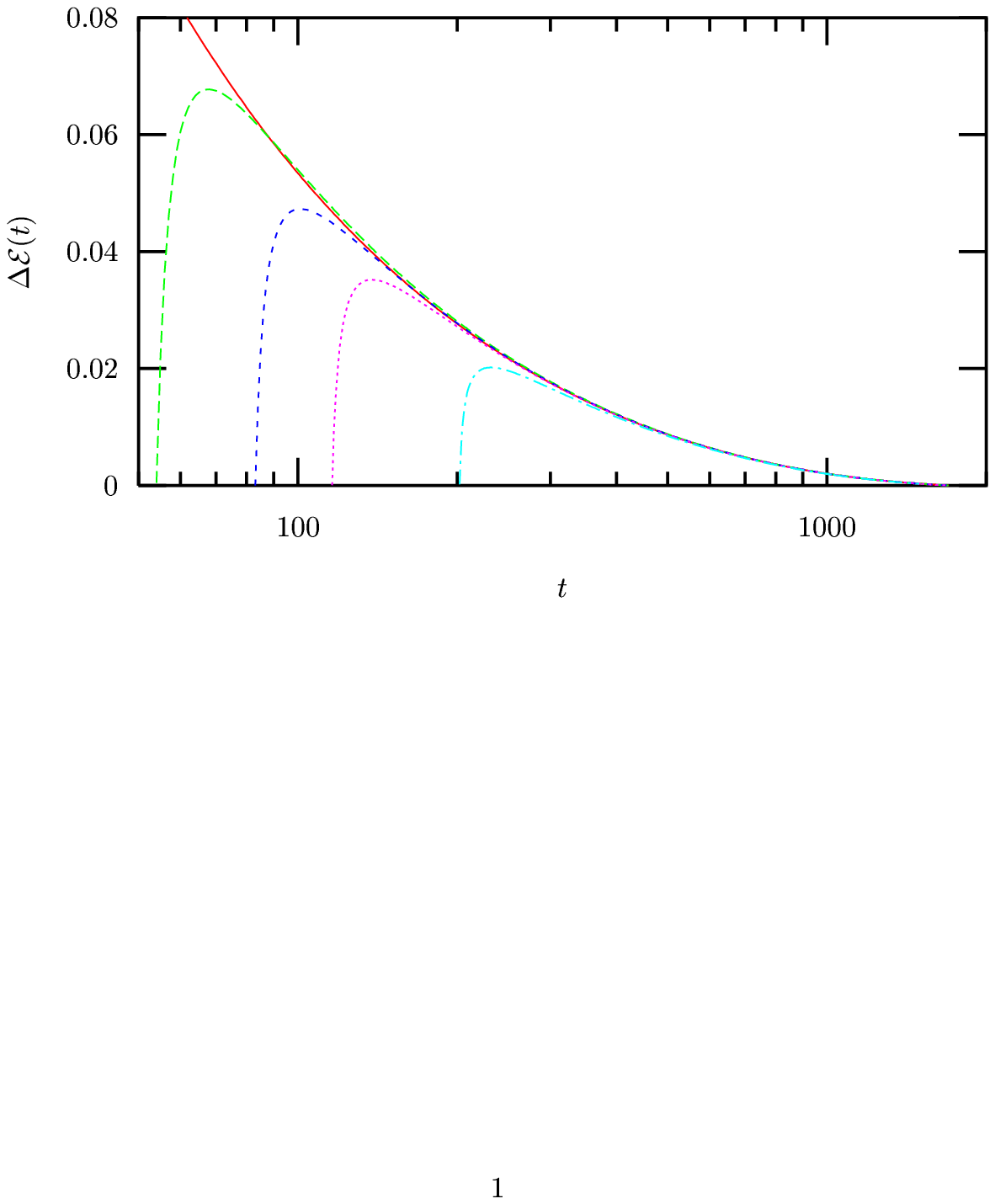}}
\setcaptionwidth{13cm} \caption{\footnotesize 
For long times the curves collapse on a master
curve under the time-shift in Eq.~(\ref{eq:shift}). With red solid line the reference curve ${\mathcal E}^{(T_2)}(t)$;
with dashed lines the curves under the effect of the temperature jump
from $T_1=0.5,0.6,0.65,0.7$ translated in time $t\to t +\Delta t$ with $\Delta t=
15, 19, 23, 31$. The final approach to zero
is exponential. }
\end{center} \end{figure}

The time-temperature dependence of the intermediate part of the
relaxation is well described with the scaling form
\begin{equation}
\Delta {\mathcal E}(t) = \Delta {\mathcal E}_K \; 
{\mathcal F}_i\left(\frac{t-t_1}{t_K-t_1}\right)
\; ,
\end{equation} 
see Fig.~\ref{fig:scaling}. This scaling law is of the class found 
in \cite{Bebo,Buhot}.

\vspace{0.25cm}
\noindent
\underline{Long times}
\vspace{0.25cm}

The long times decay of the hump, {\it i.e.} for times well beyond 
$t_K$, is exponential
\begin{equation}
\Delta {\mathcal E}(t) = a \, e^{-b t}
\; .
\end{equation}
The curves $\Delta {\mathcal E}(t)$ for different $T_1$ can be made to 
collapse by shifting time according to
\begin{equation}
t \to t + (t_K -t_1)
\label{eq:shift}
\; ,
\end{equation}
that is to say that, the curves for the systems on which a temperature 
shift was applied are in advance with respect to the reference one.

\vspace{0.35cm}

\begin{figure} \begin{center}
\resizebox{11cm}{!}{\includegraphics*[4.5cm,12cm][17.5cm,20cm]
{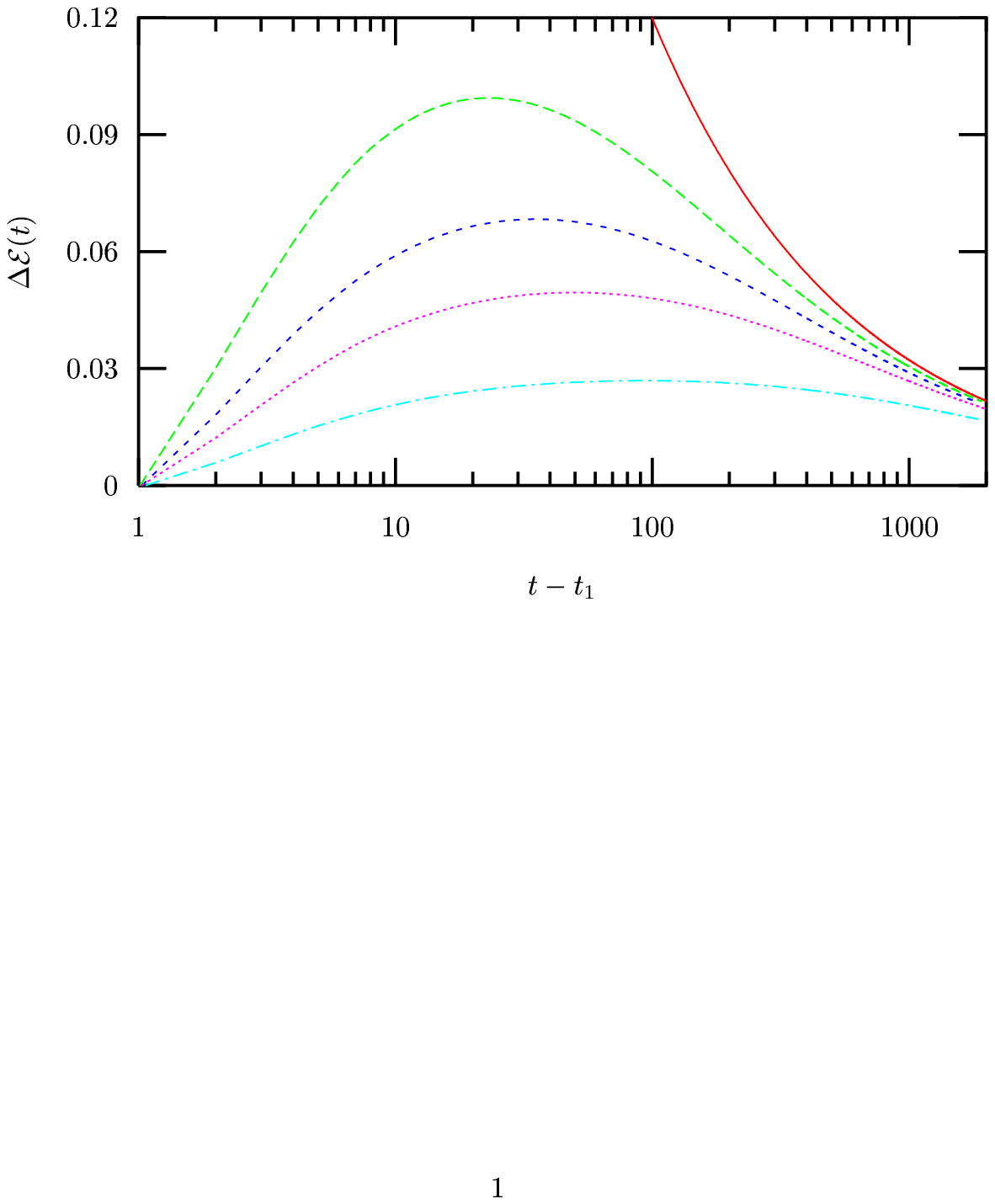}}
\setcaptionwidth{13cm} \caption{\footnotesize The hump in the glassy phase of the classical
$p$-spin model with $p=3$ using $T_1=0.3, 0.4, 0.45, 0.5$
with $t_1=80,173,330, 1080$, respectively, and $T_2=0.55$.}
\label{fig:spinglass}
\end{center} \end{figure}

Finally, we implemented the same protocol using temperatures $T_1$ and
$T_2$ that are both below the dynamic critical temperature $T_d$. As
shown in Fig.~\ref{fig:spinglass} the qualitative behaviour is the
same as on the other side of the transition point. It is interesting
to note that the curves for the perturbed system join, and
become independent of $T_1$ before reaching the reference curve (this
is similar to what observed in~\cite{Sciortino}).

In the low temperature phase the scalings discussed above are modified as
follows. In the short-times regime, the scaling function
 (\ref{eq:scaling-shortt2}) remains
valid but the scaling form is modified to
\begin{eqnarray}
\Delta {\mathcal E}(t) &=& (T_2-T_1) 
{\cal F}_s \left((t-t_1) \left(\frac{T_1}{T_2}\right)^\alpha \right)
\; ,
\\
\alpha&=& \frac{T_2-T_1}{T_d}
\; .
\end{eqnarray}
For intermediate times we found
\begin{eqnarray}
\Delta {\mathcal E}(t) &=& \Delta {\mathcal E}_K \, 
 {\mathcal F}_i\left( \left(\frac{t-t_1}{t_K-t_1} \right)^{1+\alpha} \right) 
\; ,
\end{eqnarray}
with $\mathcal{F}_i$ independent of $T_1$ and $T_2$.
At very long times, the approach to zero  is a power law,
$a t^{-b}$. The shift $t\to t+(t_K-t_1)$ is also efficient collapsing the 
data.

\subsection{Discussion}

The non-trivial content of the hump is its slow relaxation and
dependence on $T_2-T_1$. The observed behaviour has been captured by
several models already presented in the literature.  Let us discuss
these explanations and compare them to the one associated to the $p$
spin model.

First, one can compaire the behaviour observed in 
the super-cooled liquid phase with a simpler system: an overdamped 
harmonic oscillator. 
For this model a direct  calculation yields 
\begin{eqnarray}
\Delta \mathcal{E}(t) &=& \frac{1}{2} \left[k \langle x_0^2 \rangle - k_B T_1 \left(1-e^{\frac{2k}{\gamma}t_1} \right) - k_B T_2e^{\frac{2k}{\gamma}t_1} \right]e^{-2\frac{k}{\gamma}t } 
\label{eqoscilador}
\end{eqnarray}
with $k$ the harmonic constant, $\gamma$ the friction coeficient and 
$x_0$ the initial condition. The time $t_1$ is fixed by requiring that $\Delta \mathcal{E}(t_1)=0$. This condition forces  the bracket in equation (\ref{eqoscilador})  to be zero and then, there is not a forthcoming hump.  
 Clearly one needs to go beyond this model 
to get the observed 
two-temperature dependence and slow relaxation. 

A simple next step is to study a model with a distribution of relaxation 
times that depends on temperature. A simple realization is the  
$2d$ {\sc xy} model in the spin-wave approximation taking into 
account the contribution of vortices~\cite{Beho}. 
This model is given by a Gaussian 
free field (the angle of the local magnetization) 
with a $T$-dependent stiffness, $\rho(T)$. It has a slow dynamics 
characterised by the growth of a $T$-dependent correlation length $\ell_T(t)=
(\rho(T) t)^{1/2}$ and it captures the phenomenology
of the Kovacs' effect, as discussed by Berthier and Holdsworth~\cite{Beho}.

In slightly more general terms~\cite{Bebo,Bertin} the Kovacs' effect
can be rationalized in any system with a growing temperature-dependent
dynamic correlation length, that is shorter than the equilibrium one.
When one shifts the temperature to $T_2$ at $t_1$, the length scales
that are shorter than $\ell_{T_1}(t_1)$, and are hence equilibrated at
$T_1$, have to reequilibrate at $T_2$, where their equilibrium energy
is higher. The structure reached at $T_1$ has to break up and 
allow for the nucleation of new structures equilibrated at $T_2$. 
Instead, the length scales that are longer than
$\ell_{T_1}(t_1)$ are still not equilibrated at $T_1$ and they may 
continue their evolution to equilibrate now at $T_2$.  The
former processes involve shorter length-scales and should be faster
and dominate the first part of the relaxation after $t_1$ hence
leading to an energy increase. The latter processes are slower and
dominate the decay from the maximum towards the asymptotic value
${\mathcal E}_{as}(T_2)$. Within this picture, the time at which
the maximum in $\Delta {\mathcal E}$ is reached corresponds to the time
when the small length scales have equilibrated at the new temperature
$T_2$. A similar argument was put forward to explain
the overshoot observed in the time-dependent dielectric constant of
dipolar glasses after a temperature jump~\cite{dipolar}.  It is also
behind the calculation presented by Brawer on the ferromagnetic Ising
chain at very low temperature~\cite{Brawer}. 

However, it is not necessary to invoke a growing correlation length to capture 
the qualitative features of the Kovacs' effect. The main ingredient 
in the $p$-spin model that leads to this effect is the 
slow -- non-exponential -- and temperature dependent
relaxation of the linear response after a strong perturbation. A scenario
with a wide spectrum of relaxation times that depends on temperature
was used in the past to explain the Kovacs' effect~\cite{Struik}.
Here we demonstrated that, as one could have expected~\cite{Bertin}, 
the $p$-spin model or, equivalently, the  {\sc mct}
 with no equilibration assumption, has this property. 

The temperature and time dependence of the hump do 
depend on the model considered but the main qualitative features of the 
effect are shared by all of them. In the context of the spin models
related to the {\sc mct}
these will be obviously modified if one considers a mixture of $p=2$ and $p=4$ 
models (that corresponds to going from the schematic {\sc mct} to 
more refined versions).

Finally, let us discuss the relevance of the Kovacs' effect for the 
development of a thermodynamics of the nonequilibrium glassy state.

The attempts to use a thermodynamics for nonequilibrium glassy systems
are based on the introduction of effective state variables, basically
temperature and pressure. Initially, a (constant) fictive temperature
that characterizes the glassy structure was introduced by
Tool~\cite{Tool}.  As a consequence of Kovacs' experiments it was
realized that this single parameter was not enough to describe the
evolution of the glass and the fictive temperature was upgraded to be
a full history dependent function measuring the departure from
equilibrium~\cite{Kovacs,McKenna,Nara}.  Whether the fictive temperature, as 
defined in \cite{Kovacs,Tool,Nara} behaves as a
thermodynamic  temperature remains to be proven.  One can also translate
the study of the Kovacs' effect in the parking lot model by Tarjus and
Viot~\cite{Tarjus} in these terms: the second {\it time-dependent} state
variable introduced generalising Edwards' prescrition~\cite{Ed}
acts as a fictive temperature.

More recently, an effective temperature was defined using the
modification of the fluctuation -- dissipation theorem in slowly
evolving nonequilibrium systems. The interpretation of this quantity
as a {\it bonafide} temperature was discussed, and some conditions on
the physical relevance of this definition were also
given~\cite{Cukupe}. In particular, the need to have a system evolving
{\it slowly} -- characterized by a slow relaxation of one time
quantities -- to be able to associate the properties of a temperature
to the {\sc fdt} ratio was reckoned and stressed. The situation in
Kovacs' experiment goes beyond this limit: the system is strongly
perturbed and the one-time quantity under study has a non-trivial,
non-monotonic time-dependence. The Fluctuation Dissipation Relation
also depends on time in a non-trivial manner~\cite{Buhot2}. 
Thus one cannot assert
that it leads to a {\it bonafide} temperature and use it to construct
a simple extension of thermodynamics to describe the subsequent
behaviour of the system.

A similar conclusion, though expressed in terms of the potential
energy landscape ({\sc pel}) scenario was reached by Mossa and 
Sciortino~\cite{Sciortino}. These authors
showed, with molecular dynamic simulations of the fragile glass former
{\sc otp}, that two systems in identical thermodynamic conditions
(same values of $T,V,P$) can be in very different regions of their
potential energy landscape if one of them has been strongly
perturbed.  Since the 
strongly perturbed system wanders in a region of the {\sc pel} that is 
never sampled in equilibrium, its configuration cannot be associated
to that of an equilibrated liquid at a different temperature.
The region of the 
{\sc pel} sampled allows for a definition of 
a microcanonical temperature only when the variation of the external 
conditions (temperature in this case) is small.
Mossa and Sciortino arrived at this conclusion by comparing the properties
of the inherent structures (closest local minimum in the {\sc pel}) 
visited during aging after a temperature jump of
large magnitude to those sampled in equilibrium.

\section{Quantum model}
\label{sect:quantum}

When quantum fluctuations are switched on one has to deal with the full
Schwinger-Dyson equations (\ref{schwingerR})-(\ref{kernelD}). As 
explained in~\cite{Cugrsa}
the parameter that plays the r\^ole of the transverse field in trully quantum
spin models is here the inverse of the mass $M$. 
Indeed, this model undergoes a phase transition in the 
$(T,\Gamma)$ plane with $\Gamma\equiv \hbar^2/(M{\tilde J})$
from an equilibrium paramagnetic phase
to a non-equilibrium glassy one. In order to test the memory effect 
in the quantum problem, we then apply the Kovacs' protocol using $M^{-1}$ 
as the control parameter and we follow the evolution of the averaged potential 
energy density ${\mathcal E}(t)$.

\begin{figure}[t] \begin{center}
\resizebox{11cm}{!}{\includegraphics*[4.5cm,12cm][17.5cm,20cm]
{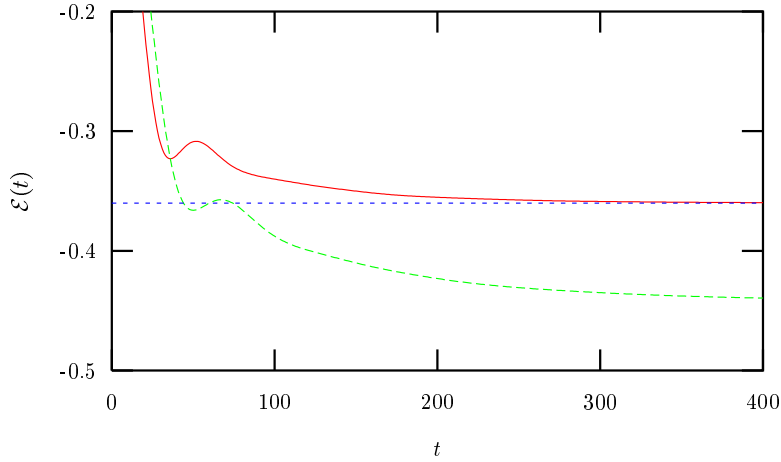}}
\setcaptionwidth{13cm} \caption{\footnotesize Energy-density decay in the quantum model with $p=3$
at constant mass and temperature. 
Solid line $M=0.5$ and dashed line $M=0.8$. In both cases $T=0.75$.}
\label{fig:quantum-E}
\end{center} \end{figure}

In the quantum problem the asymptotic value of the potential energy density 
depends on $M$. (In the classical limit it does not.) The 
relaxation of the potential energy density
at constant $M$ has (damped) oscillations whose magnitude depend on
the parameter $M$.
For large values of the mass the oscillations have a
sufficiently large amplitude such that the asymptotic value falls
within the oscillation, {\it i.e.}  ${\mathcal E}(t) < {\cal
E}_{as}(T_2,\Gamma_2) $ for some finite times $t$.  In the following
we choose a value of $M_2$ such that the system is close to the
paramagnetic -- glass transition and for which 
${\mathcal E}_{as}(T_2,\Gamma_2) 
< {\mathcal E}(t)$ for all finite time $t$, see Fig.~\ref{fig:quantum-E}.
Another feature to signal is that the oscillation in the 
energy-density decay may be such that
there is more than one value of $t_1$ for 
which ${\mathcal E}^{M_1}(t_1^+) = {\cal
E}^{M_2}_{as}$, see the curve drawn with a dashed line in 
Fig.~\ref{fig:quantum-E} for which $M=0.8$.

\begin{figure} \begin{center}
\resizebox{11cm}{!}{\includegraphics*[4.5cm,12cm][17.5cm,20cm]
{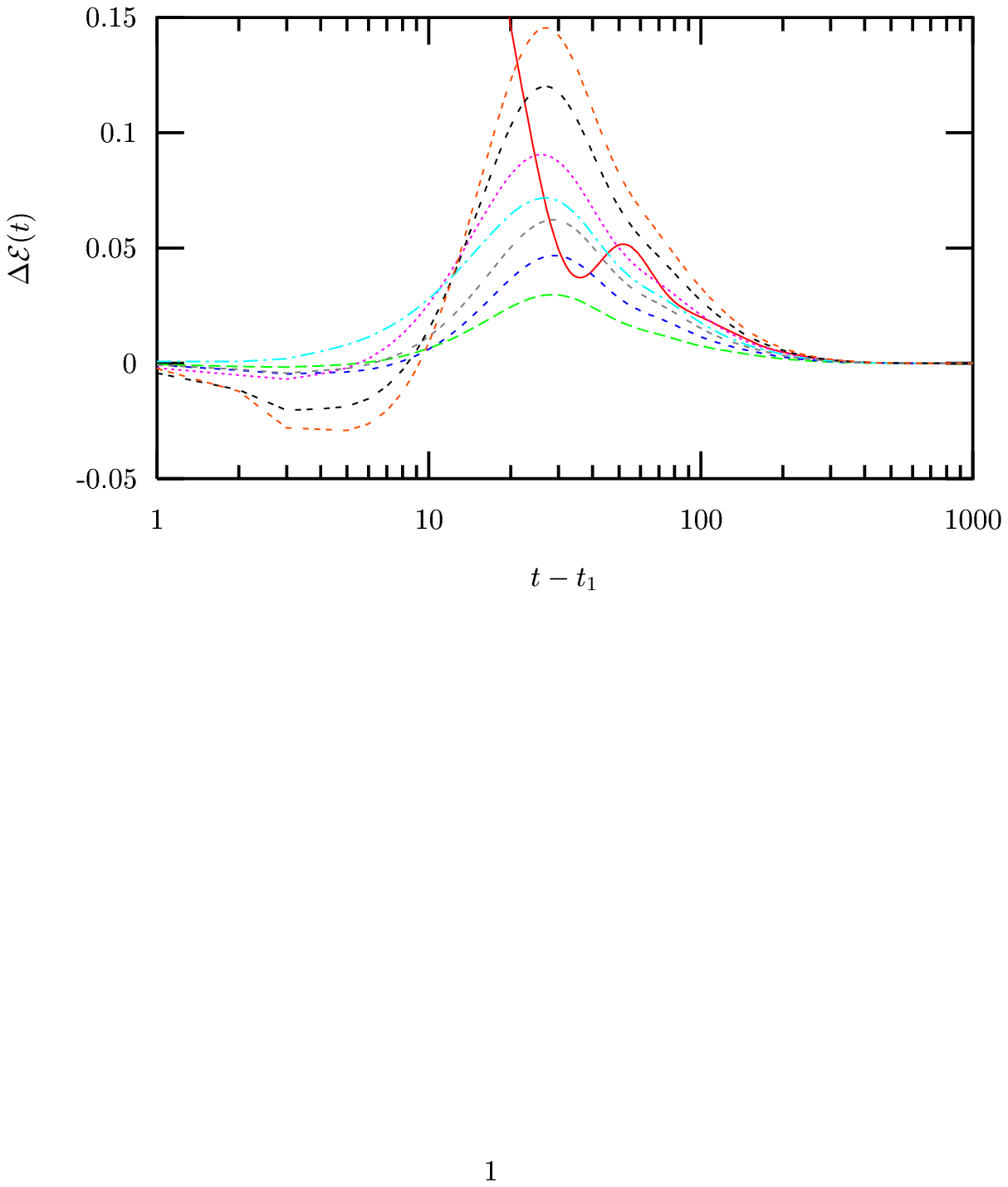}}
\setcaptionwidth{13cm} \caption{\footnotesize Hump in the quantum $p=3$ model at $T=0.75$. The reference 
curve, solid red line, corresponds to $M_2=0.5$. The modified curves 
have been obtained using, from bottom to top at the maximum, 
$M_1=0.6$, $t_1=95$ (dashed green);
$M_1=0.7$, $t_1=82$ (dashed blue);
 $M_1=0.8$, $t_1=76$ (dashed gray);
 $M_1=0.8$, $t_1=62$ (dashed cyan); 
$M_1=0.8$, $t_1=46$ (dashed magenta);
 $M_1=1$, $t_1=47$ (dashed black) and $M_1=1.2$, $t_1=49$ (dashed
orange).}
\label{fig:quantum}
\vspace{0.35cm}
\resizebox{11cm}{!}{\includegraphics*[4.5cm,12cm][17.5cm,20cm]
{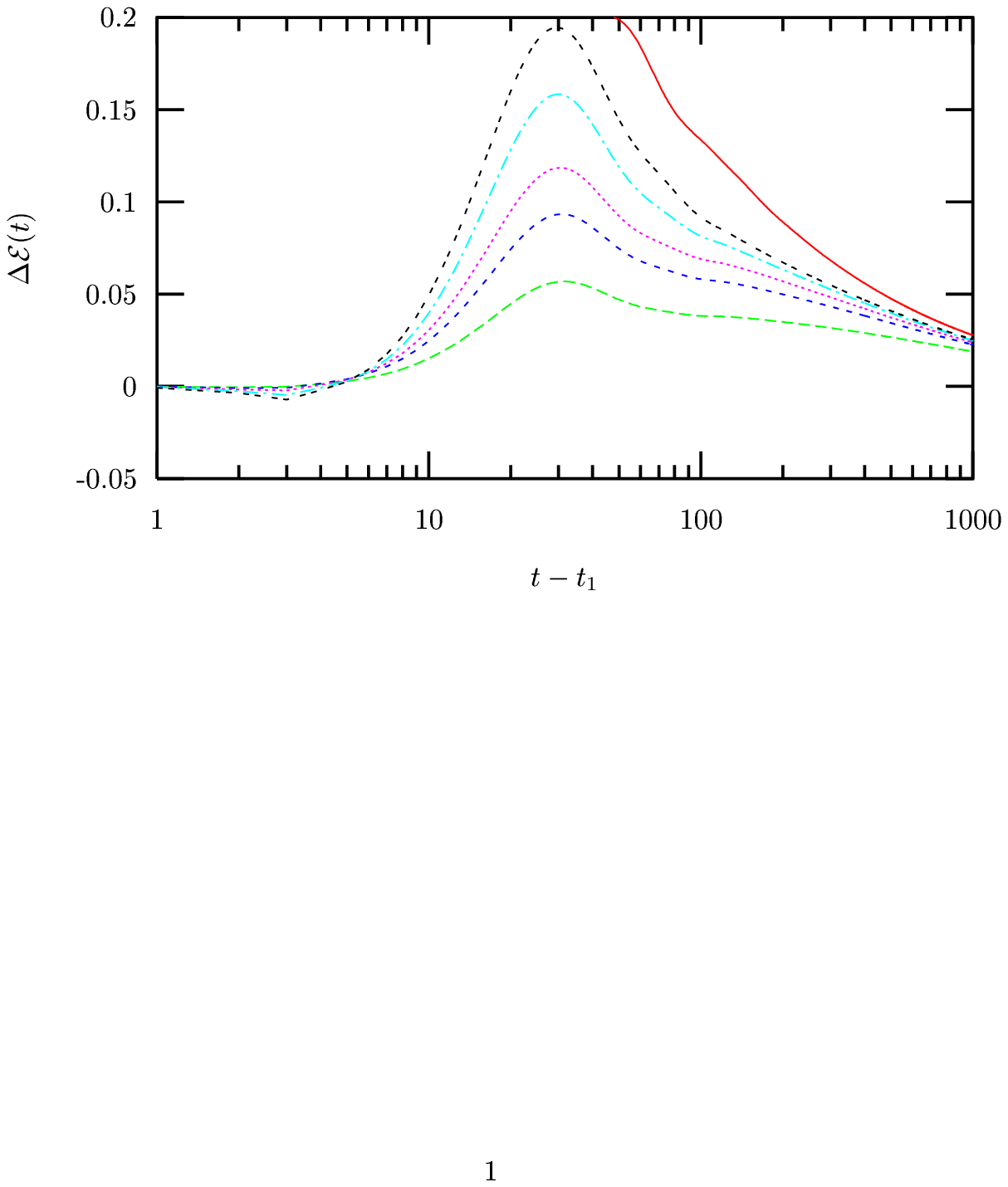}}
\setcaptionwidth{13cm} \caption{\footnotesize The hump as a function of the time-difference $t-t_1$ in the glassy phase of the quantum  $p=3$ model using, from bottom to top, $M_1=0.6, t_1=539$ (dashed green), 
$M_1=0.7, t_1=274$ (dashed blue),
$M_1=0.8, t_1=204$ (dashed magenta),
$M_1=1, t_1=163$ (dashed cyan),
$M_1=1.2, t_1=150$ (dashed black), and the     
reference curve at constant $M_2=0.5$ (solid red). The temperature is $T=0.2$ in all 
cases.}
\label{fig:quantum-glassy}
\end{center} \end{figure}

Figure~\ref{fig:quantum} shows the result of applying Kovacs'
protocol to a system at $T=0.75$ with the reference value of the mass,
$M_2=0.5$.  We use four values of $M_1$, $M_1=0.6$, $0.8$, $1$ and $1.2$ that
satisfy $M_1 > M_2$, {\it i.e.} $\Gamma_1 < \Gamma_2$.  For $M_1=0.6$, $1$ and 
$1.2$ we find a unique $t_1$ satisfying ${\mathcal E}^{M_1}(t_1) =
{\mathcal E}^{M_2}_{as}$. 
For $M_1=0.8$ instead three solutions to this equation exist
as shown in Fig.~\ref{fig:quantum-E}.  Displayed in
Fig.~\ref{fig:quantum} are the reference energy density ${\cal
E}^{(T_2,\Gamma_2)}(t)$ (solid red line) and the hump in the energy
densities obtained by shifting the mass.

The first thing to note in the figure is that the curves depend on 
the value of $M_1$, similarly to what happened in the classical 
case with $T_1$: the larger the difference in the masses 
(or in the quantum parameter $\Gamma$), the 
more pronounced the effect. There is also a weak decreasing dependence of the 
position of the maximum with $M_2-M_1$ (and $\Gamma_1-\Gamma_2$). 

A second feature to remark is that the curves
following the jump in $M$ go above the reference curve ({\it cfr.} 
the classical problem where the approach to the reference curve always
occurs from below) and also might have a negative initial part.  

In the case in which there is more than one $t_1$, 
the form of the hump depends on the value chosen.
  In particular, there is no negative
part in the hump if we take  $t_1$ such that ${\cal
E}^{(M_1\to M_2)}(t_1)$ is growing with $t$.

\vspace{0.35cm}

Finally, in the glassy phase ($T=0.2$ and $M=0.5$) the oscillation in
${\mathcal E}(t)$ is almost completely damped.  The hump has a very
similar behaviour to the one found after a shift in
$T$. Figure~\ref{fig:quantum-glassy} shows the hump for several values
of the pairs $(M_1, t_1)$ given in the caption.  Also in this glassy
case, the height of the hump increases with the difference $M_2-M_1$
and, simultaneously, the position of the maximum has a very smooth drift
towards smaller values of $t-t_1$.

Thus, as far as the Kovacs' effect is concerned, we see that the quantum 
problem also shows a non-trivial dependence on the parameter 
$M_1$ ($\Gamma_1$) and a slow relaxation after the perturbation.
 
\section{Conclusions}
\label{sect:conc}

We conclude that models with no spatial structure, like the 
$p$ spin spherical model that is intimately related to the schematic 
mode-coupling theory, can reproduce non-trivial memory effects when 
their non-equilibrium dynamics is studied. Similarly to what observed
when reproducing the hole-burning protocol~\cite{JLuis}, we found here 
that the Kovacs' memory effect is captured by this model.
In this sense, the Kovacs' experiment is not able to prove the 
existence of a growing correlation length in glassy systems.
Assuming that a length scale exists one could, however, compare the outcome 
of this and other experiments to what can be derived from 
a domain-growth like picture for glassy dynamics.

The reason for having these non-trivial long-memory effects in these
fully - connected spin models is that close to their dynamic critical
temperature (and below it) their  response function after
 strong perturbation has been applied is not given by a simple
exponential relaxation.  The slow decay of the reponse implies
that the effect of non-linear perturbations takes very long to
disappear. This is encoded in the Schwinger-Dyson equations 
which describe the dynamical evolution of the system.

\vspace{1cm}
\noindent
\underline{Acknowledegments}
We acknowledge financial support from the an Ecos-Sud travel grant
and the ACI project "Optimisation algorithms and quantum disordered
systems". LFC is research associate at ICTP Trieste and
a fellow of the J. S. Guggenheim Foundation.
G. S. L. is supported by CONICET. This research was supported in part
by SECYT PICS 03-11609, PICS 03-05179 and 
by the National Science Foundation under Grant No. PHYS99-07949.
We thank J-P Bouchaud and D. R. Grempel for very useful discussions.

\end{document}